# Transient laser-induced periodic surface structures revealed by time-resolved EUV diffuse scattering

D. Ksenzov,[1,*,†] F. Capotondi,[2,*] A. A. Maznev,[3,*] F. Bencivenga,[2] D. Engel,[4] D. Fausti,[2,5,6] L. Foglia,[2] R. Gruber,[7] N. Jaouen,[8,9] M. Kläui,[7,10] I. Nikolov,[2] M. Pancaldi,[2,8] E. Pedersoli,[2] B. Pfau,[4] C. Gutt[1]

[1]Universität Siegen, Walter-Flex-Straße 3, 57072 Siegen, Germany.

[2]Elettra Sincrotrone Trieste, Strada Statale 14, km 163.5, 34149 Basovizza, TS, Italy.

[3]Department of Chemistry, Massachusetts Institute of Technology Cambridge, Massachusetts 02139, USA.

[4]Max Born Institute, Max-Born-Straße 2A, 12489 Berlin, Germany.

[5]Department of Physics, Università degli Studi di Trieste, 34127, Trieste, Italy

[6]Department of Physics, University of Erlangen-Nürnberg, 91058, Erlangen, Germany

[7]Institut für Physik, Johannes Gutenberg-Universität Mainz, 55099, Mainz, Germany

[8]Department of Molecular Sciences and Nanosystems, Ca' Foscari University of Venice, Venice, Italy

[9]Synchrotron SOLEIL, Saint-Aubin, Boite Postale 48, 91192, Gif-sur-Yvette Cedex, France.

[10]Graduate School of Excellence Materials Science in Mainz, 55128, Mainz, Germany

* These authors contributed equally to this work.

† Contact author: dmitriy.ksenzov@uni-siegen.de

**Keywords:** Laser-induced periodic surface structures (LIPSS), low-spatial-frequency LIPSS (LSFL), ultrafast surface dynamics, low-fluence laser excitation, time-resolved x-ray scattering, transient surface modulation.

**Abstract**

The formation of permanent laser-induced periodic surface structures (LIPSS) on solid surfaces under impulsive laser irradiation above the damage threshold has been subject of extensive research. We demonstrate the formation of *transient* surface displacement patterns under femtosecond laser irradiation at fluences well below this threshold. Time-resolved extreme ultraviolet scattering measurements reveal distinct reciprocal-space features similar to those observed for permanent LIPSS but dissipating on the hundreds-of-picoseconds time scale. We show that the transient surface displacement patterns responsible for these features are produced via thermal expansion by the spatial modulation of absorbed laser intensity caused by scattering of the laser radiation by surface roughness and present a model accounting for the experimental observations. We suggest that our experiment revealed a universal phenomenon that will be observed on any strongly absorbing material under ultrafast laser irradiation.

**Introduction**

Laser-induced periodic surface structures (LIPSS) often spontaneously emerge on solid surfaces irradiated by intense laser pulses.[1–4] These structures are of significant interest for various applications, including surface engineering on micro- and nanoscale and the development of optical devices.[5–7] The LIPSS formation occurs under high-fluence laser irradiation, resulting in permanent changes to the sample surface through processes such as melting, ablation, or mass reorganization. In strongly absorbing materials irradiated by multiple laser pulses slightly above the damage threshold, the most commonly observed structures known as low-spatial-frequency LIPSS exhibit a periodicity on the order of the laser wavelength and are typically oriented perpendicular to the polarization direction of the incident light.[4,8] The origin of these structures is the interference of the laser radiation scattered from surface roughness with the incident laser beam leading to spatially modulated energy deposition described by the theory developed by J. E. Sipe and others.[1,4,9] The origin of some other kinds of LIPSS, especially those with a periodicity much smaller than the laser wavelength, is still not clearly understood, and in general, the LIPSS formation continues to be a subject of intense research.[10–12]

Time-resolved diffuse scattering of femtosecond extreme ultraviolet (EUV) pulses produced by free electron lasers has recently emerged as a powerful tool for studying ultrafast nanoscale phenomena, such as the evolution of magnetic textures under femtosecond laser excitation.[13,14] Investigating the LIPSS formation in the Fourier space is a promising application of this technique.[12]

In this work, we study time-resolved diffuse extreme ultraviolet scattering (DES) following femtosecond laser irradiation of metal films well under the damage threshold. In the scattering patterns, we observe characteristic arcs similar to those produced by the "classic" low-spatial-frequency LIPSS, exhibiting the same dependence on the laser wavelength and polarization. These features are transient, appearing within a few picoseconds after the laser excitation and fading away on a timescale of hundreds of picoseconds. Interestingly, the most prominent features correspond to a decrease in the scattering intensity compared to static scattering. We suggest that the origin of the observed scattering patterns is a surface displacement modulation produced by non-uniform thermal expansion resulting from the same scattering and interference effects that are responsible for the formation of LIPSS at higher laser fluences. We demonstrate that our model, in conjunction with Sipe's theory, accurately describes the observed scattering features and that their decay time is consistent with a mechanism based on thermal diffusion. Finally, we discuss the implications of transient LIPSS-like structures and their relationship with concurrently observed circular fringe patterns recently reported in Ref.[15]

**Experimental**

The experiment was performed at the DiProI end-station of the FERMI Free-Electron Laser (FEL) facility in Trieste, Italy[16]. The experimental configuration for the time-resolved soft X-ray scattering experiments is shown schematically in Figure 1(a). The sample surface was positioned at an incidence angle of approximately 45° relative to the incoming FEL pulses. A nearly collinear optical pump pulse with linear polarization, either at the fundamental laser wavelength ($\lambda_L$ = 790 nm, p-polarized, pulse duration ∼ 60 fs, sample-to-detector distance 140 mm) or second harmonic ($\lambda_L$ = 395 nm, s-polarized, pulse duration ∼ 90 fs, sample-to-detector distance 55 mm), was overlapped with the FEL probe pulse on the sample surface. The optical pump delivered a fluence in the range of 10–30 mJ/cm², with a spot size of 300 × 250 µm² at 790 nm and 570 × 540 µm² at 395 nm.

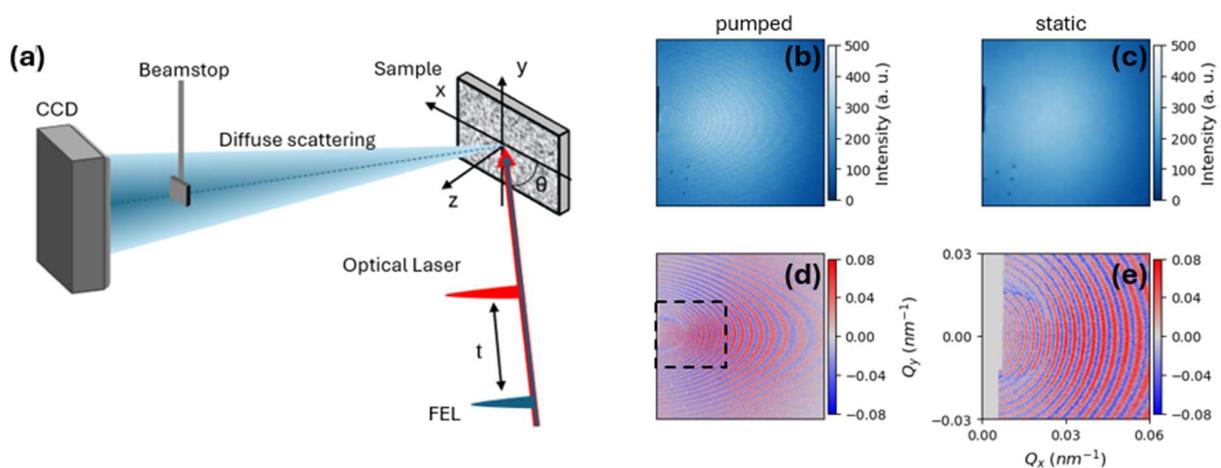

**Figure 1. (a)** Schematic of the experimental setup. An optical laser pulse is used as a pump, while an FEL pulse acts as a probe. The DES signal is collected by a CCD detector positioned at ~45° relative to the sample surface. Representative DES intensity maps are shown for a [Ta(2.4 nm)/Pt(2.4 nm)]×4 multilayer stack grown on top of a 100-nm-thick SiN film on a silicon substrate excited at 395 nm: **(b)** DES pattern recorded 900 ps after the optical excitation, **(c)** static scattering pattern without the optical pump, **(d)** the differential image obtained by subtracting (c) from (b) and normalized to the mean intensity of the static pattern, and **(e)** the transformed differential image represented in terms of the in-plane wave vector components $Q_x$ and $Q_y$, showing only the region indicated by a dashed rectangle in (d).

The FEL probe pulses had a duration of about 60 fs and a repetition rate of 50 Hz. The probe wavelength was 20.8 nm for measurements with 790 nm excitation and 17.8 nm for measurements with 395 nm excitation. The FEL beam was focused to a spot size of 150 × 120 µm² with a fluence of approximately 1.5 mJ/cm². The time delay $t$ between the optical pump and the FEL probe was varied up to 1 ns using an optical delay line. The scattered EUV radiation was measured in reflection geometry using a two-dimensional charge-coupled device (CCD) detector. The specularly reflected FEL beam was blocked by a beamstop. At each delay time, scattering patterns were recorded by integrating over 600 laser shots for 395 nm excitation and 1,000 shots for 790 nm excitation and normalized to the incident FEL flux (a representative CCD image is shown in Fig. 1b). A static scattering image (Fig. 1c) obtained with the pump laser beam blocked was subtracted, and the resulting differential scattering pattern (Fig. 1d) was normalized to the mean intensity of the static image to allow comparison between different samples. Finally, following the methodology described by Capotondi *et al.*[15], an isomorphic coordinate transformation was applied to convert the CCD image into the scattered intensity distribution in the reciprocal space $\Delta I(Q_x, Q_y, t)$ where $Q_x$ and $Q_y$ denote the in-plane scattering wave vector components parallel and perpendicular to the projection of the incident beam on the sample surface (Fig. 1e).

Control measurements indicated that while the patterns on the detector depend on the FEL wavelength, the transformed images in the reciprocal space are independent of the FEL wavelength, see Supplementary Figure S1.

**Results**

Figure 2 (a–d) shows differential DES images at different time delays from a [Ta(2.4 nm)/Pt(2.4 nm)]×4 multilayer stack grown on top of a 100-nm-thick SiN film on a silicon substrate excited at 395 nm, while Fig. 2 (e–h) shows a series of images from a [Ir(6.6 nm)/CoFeB(0.7 nm)/MgO(2.0 nm)]×15 multilayer stack grown on a Si/SiO2 substrate under 790 nm excitation.

At positive delays, distinct arc-like features indicated by grey arrows in Figures 2(b) and 2(f), appear at locations predicted by Sipe's LIPSS theory[1] which deals with the interference of the optical field scattered from surface roughness with the incident pump beam. Sipe *at al.* showed that the resulting absorbed intensity pattern has sharp peaks in the reciprocal space occurring at the wave vectors satisfying

$$Q_x = k_0 \sin \varphi \, (1 + \cos \theta)$$
$$Q_y = k_0 \sin \varphi \tag{1}$$

where $k_0 = 2\pi/\lambda_L$ is the optical wave vector, $\varphi$ is the azimuthal angle and $\theta = 45°$ is the incidence angle. Eq. (1) describes circles of radius $k_0$ centered at $Q_x = k_0 \cos \theta$, $Q_y = 0$. As one can see in Fig. 2, the observed arcs perfectly match these circles. Furthermore, the arcs are orthogonal to the projection of the polarization on the surface, which also agrees with Sipe's theory. The locations of the arcs correspond to spatial periods about 350 nm for the 395 nm excitation and 460 nm for the 790 nm excitation. Interestingly, the observed LIPSS-like features correspond to a negative differential scattering intensity, i.e. the scattering intensity is reduced compared to the static scattering background.

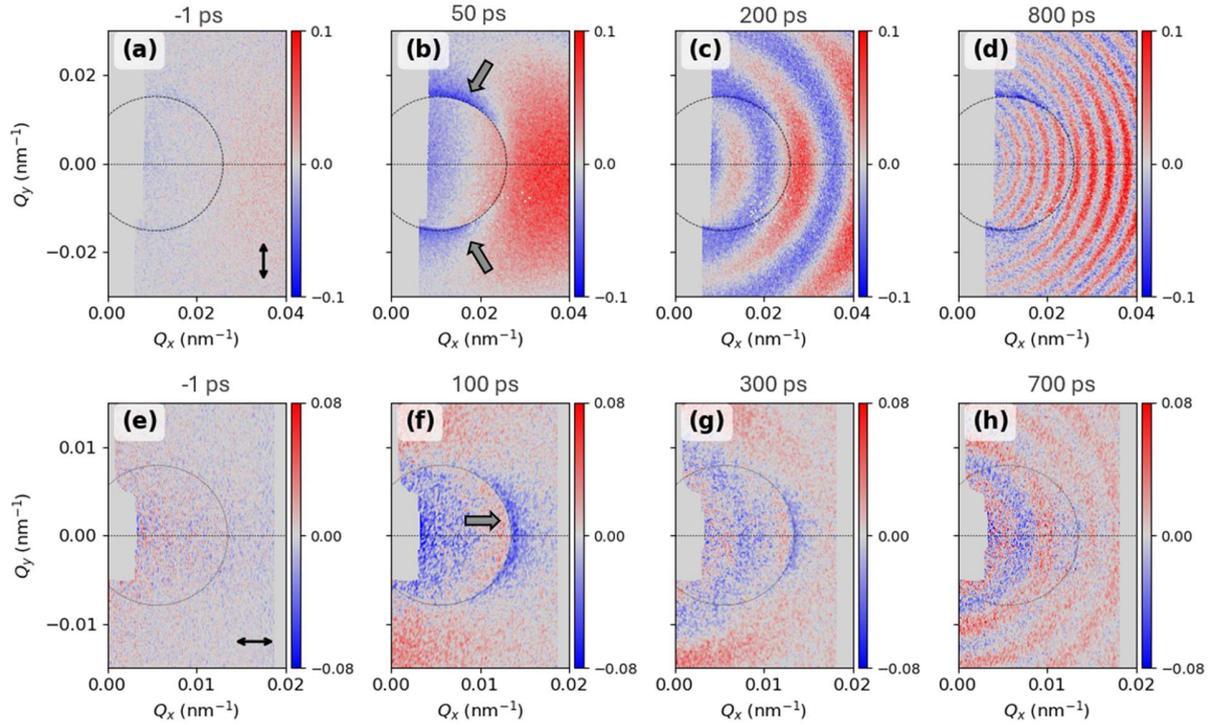

**Figure 2.** Temporal evolution of the differential EUV diffuse scattering patterns for **(a–d)** the Ta/Pt multilayer sample excited by 395 nm s-polarized pump pulses and **(e–h)** Ir/CoFeB/MgO sample excited at 790 nm and p-polarization at different pump-probe time delays indicated on the graphs. Gray arrows in panels (b) and (f) mark the transient LIPSS-like arcs. Dotted circles show the theoretical location of LIPSS arcs in the reciprocal space according to Eq. (1).

The LIPSS-like arcs dissipate on the hundreds of picoseconds time scale. Concurrently, a system of concentric circular fringes centered at $Q = 0$ develops; these fringes investigated in the recent report[15] are produced by spatially random but temporally coherent surface acoustic waves (SAWs) launched at t=0 and covering a wide wave vector range. We verified that after the pump laser was turned off, no patterns were observed, i.e. no permanent structures were formed on the sample surface. Similar transient LIPSS-like features were also observed on several single metal film samples (Pt, Co, Ta, Ti), see Supplementary Figure S2.

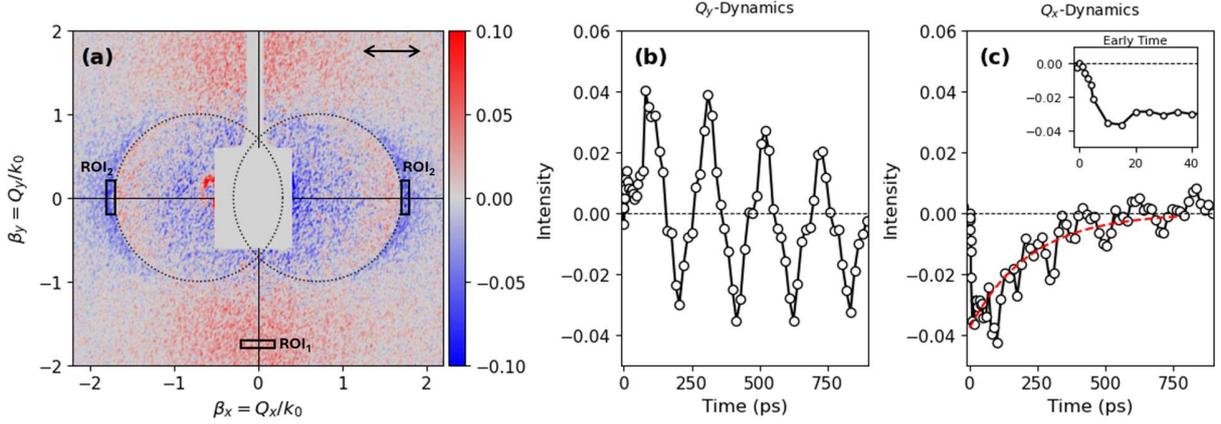

**Figure 3.** Time-domain analysis of the transient surface response. **(a)** Regions ROI$_1$ and ROI$_2$ shown on the DES image from the Ir/CoFeB/MgO sample taken at $t = 100$ ps. Dotted circles are given by Eq. (1). **(b)** Time evolution of the signal averaged within ROI$_1$. **(c)** Time evolution of the signal within ROI$_2$. The red curve represents a fit with Eq. (2). The inset shows the initial fast dynamics.

To analyze the temporal evolution of the scattering signal, we defined small rectangular regions, ROI$_1$ and ROI$_2$, in the scattering patterns from the [Ir/CoFeB/MgO]×15 sample, as shown in Fig. 3(a). (Note that Fig. 3(a) shows the full scattering image collected by the CCD, whereas only half of the images are shown in Fig. 2(e–h) for compactness).

Region ROI$_1$ is defined by $\beta_x \in [-0.2, 0.2]$ and $\beta_y \in [-1.8, -1.7]$, whereas ROI$_2$ includes the symmetric regions $\beta_x \in [\pm 1.7, \pm 1.8]$ and $\beta_y \in [-0.2, 0.2]$, where $\boldsymbol{\beta} = \boldsymbol{Q}/k_0$ is dimensionless scattering wave vector. While the magnitude $|\boldsymbol{\beta}|$ is the same in both regions, ROI1 lies outside of the transient LIPSS-like feature, while ROI2 is located at the center of the LIPSS-like arc.

Figures 3(b) and (c) show the time evolution of the scattering intensity within regions ROI1 and ROI2, respectively. ROI1 yields an oscillatory signal produced by SAWs, in agreement with Ref. [15]. In contrast, ROI2 yields dynamics with a sharp rise and slow decay. SAW oscillations are also present but are less prominent and shifted in phase compared to ROI1. The red curve in Figure 3(c) shows a fit with a single-exponential model,

$$\Delta I(t) = A_1 e^{-\frac{t}{\tau_d}} \qquad (2)$$

where $\tau_d = 220$ ps is the decay time and $A_1 < 0$ represents the amplitude of the transient change in scattering intensity. The inset in Fig. 3(c) shows the initial dynamics: the signal reaches ~60% of its maximum amplitude within 6 ps. The observed dynamics resemble those seen in transient-grating experiments where the sample is excited by a spatially periodic optical intensity pattern.[17–19]

**Theoretical model**

We rely on the electromagnetic scattering theory developed by Sipe and others[1,9] that deals with the interference of the pump light scattered by surface roughness with the incident pump beam resulting in a spatial modulation of the absorbed energy density per unit area $E_{abs}$. Following Ref. [8], we start by considering a sinusoidal surface roughness profile $z(\boldsymbol{r}) = h \cos \boldsymbol{Q}\boldsymbol{r}$, in which case $E_{abs}$ is given by

$$E_{abs}(\mathbf{r}) = E_0[1 + hk_0(A \cos \boldsymbol{Q}\mathbf{r} + B \sin \boldsymbol{Q}\mathbf{r})], \qquad (3)$$

where $\boldsymbol{Q}$ and $h$ are the wave vector and amplitude of the surface corrugation, $\mathbf{r}$ is the in-plane coordinate on the sample surface, $E_0$ is the absorbed energy density for a perfectly flat surface, and *A, B* are dimensionless coefficients encoding the electromagnetic response of the system which can be readily obtained from Refs.[9,20]. The absorbed energy pattern has the same spatial period as the surface roughness but is generally shifted in phase relative to it, unless the pump beam is normal to the surface. The sinusoidal spatial modulation of absorbed energy acts similarly to the transient grating excitation[18,19] in which the sinusoidal modulation is created by the interference of two pump beams. It is well known[19,21] that such an excitation creates, via thermal expansion, a sinusoidal surface displacement profile.

For the sake of simplicity, we will ignore the layered structure of the film stack and assume that the sample is a uniform half-space. Assuming that the laser radiation is absorbed within a shallow skin depth which is much smaller than the period $\Lambda = 2\pi/Q$, the laser-induced temperature rise and subsequent thermal expansion result in a sinusoidal modulation of the vertical surface displacement $u(\mathbf{r})$ replicating the spatial profile of $E_{abs}$[21],

$$u(x) = h\varepsilon(A \cos \mathbf{Qr} + B \sin \mathbf{Qr}), \tag{4}$$

with $\varepsilon$ is a dimensionless factor given by

$$\varepsilon = \frac{2(1+\nu)k_0\alpha E_0}{\rho c}, \tag{5}$$

where $\alpha$ is the linear thermal expansion coefficient, $\rho$ is the density, *c* is the specific heat, and $\nu$ is Poisson's ratio. The total surface corrugation, combining the static roughness and the laser-induced displacement, becomes

$$z_{tot}(\mathbf{r}) = z(\mathbf{r}) + u(\mathbf{r}) = h[(1 + \varepsilon A)\cos \mathbf{Qr} + \varepsilon B \sin \mathbf{Qr}]. \tag{6}$$

The scattering intensity from a small amplitude sinusoidal surface profile $a \cos qx + b \sin qx$ is proportional to the square of the amplitude[22], i.e., $a^2 + b^2$. Consequently, the scattering intensity at the wave vector $\mathbf{Q}$ will be given by

$$I_{pumped} = Ch^2[(1+\varepsilon A)^2 + (\varepsilon B)^2], \qquad (7)$$

where *C* is a proportionality constant, whereas the static scattering intensity in the absence of the laser pump will be given by $I_{static} = Ch^2$. We know from the experiment that the laser-induced scattering intensity variations are small compared to the static scattering; therefore, laser-induced displacement is small compared to the static roughness, i.e., $\varepsilon A, \varepsilon B \ll 1$. Consequently, the normalized differential scattering intensity can be linearized as follows

$$\frac{\Delta I}{I_{static}} \approx 2\varepsilon A. \qquad (8)$$

The sign and magnitude of the laser-induced scattering signal are governed by the coefficient *A*. A positive value of *A* means that the absorbed laser energy modulation is in phase with the static surface corrugation; as a result, the laser-induced displacement adds to the static displacement thus increasing the total scattering intensity. When *A* is negative, the laser-induced displacement is out-of-phase with the static corrugation, and the scattering intensity is reduced. In other words, EUV light scattered by the laser-induced displacement interferes constructively or destructively with the static scattering depending on the sign of *A*; static scattering acts as a local oscillator amplifying the signal via heterodyne detection.[23] The $\varepsilon B$ term contributes only through $(\varepsilon B)^2$ in Eq. (7) and is therefore negligible in the differential scattering signal.

While we started our analysis by assuming a sinusoidal surface roughness profile, an arbitrary profile *z(r)* can be decomposed into sinusoidal components. Each Fourier-component of the surface roughness effectively results in a sinusoidal modulation of the absorbed optical energy acting as a transient grating excitation. Since the scattering intensity is detected in the reciprocal space, the signal received by a given pixel on the CCD is equivalent to the transient grating signal at the corresponding wave vector ***Q***. Consequently, Eq. (8) is valid for an arbitrary surface roughness profile, and the distribution of the normalized differential scattering intensity in the reciprocal space is determined solely by $A(\mathbf{Q})$. It is assumed, of course, that the surface roughness is small compared to the EUV probe wavelength. The rms roughness of our samples was about 1-2 nm.

Figure 4 show the comparison between the modulation coefficient $A(\mathbf{Q})$ for an Ir film, obtained using the electromagnetic interference model of LIPSS[9] (see Supplementary Materials, Section S3) and the experimental differential scattering profile along the $Q_x$ axis in the reciprocal space. The narrow positive peak predicted by the model is not resolved in the experiment due to the limited reciprocal-space resolution caused by averaging over the probe beam footprint. The measured profile displays a broad negative peak predicted by the model. The qualitative agreement between the calculated and experimental profiles indicates that the observed transient scattering anisotropy arises from the same electromagnetic-interference mechanism that underlies the LIPSS formation.

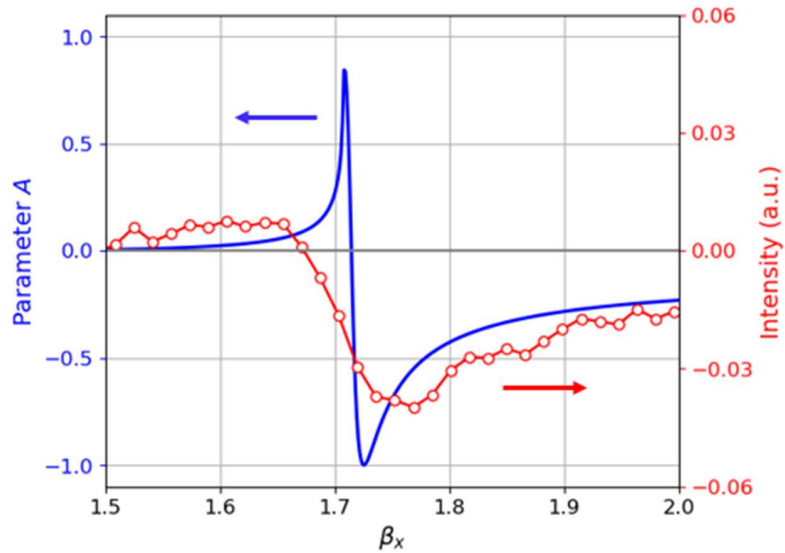

**Figure 4.** Comparison between the calculated one-dimensional profile $A(\beta_x)$ for the Ir film (blue line), obtained using the electromagnetic interference model of LIPSS[9] under p-polarized excitation and the experimental scattering profile measured along the same direction (red circles).

Let us now consider the dynamics of the LIPSS-like feature. The surface displacement does not appear instantaneously: for a bulk sample or a film whose thickness exceeds the optical skin depth $\zeta$, the time scale of the initial thermal expansion is given by $\zeta/v_l$, where $v_l$ is the longitudinal speed of sound.[24] For our multilayer stack the situation is more complicated, but we could still use $\zeta/v_l \sim$ 3 ps for Ir as a rough estimate. This is reasonably consistent with the experimental value 6 ps, considering that the rise time of the signal is not well sampled and the experimental value may have a significant uncertainty.

Furthermore, surface displacement due to the spatially sinusoidal temperature modulation will not remain static; rather, it will be washed away by thermal diffusion. Assuming that for our multilayer stack in-plane thermal transport dominates, the "thermal grating" will decay exponentially with the decay time given by[18]

$$\tau_d = \frac{1}{D_T Q^2}, \qquad (9)$$

where $D_T$ is the thermal diffusivity. Using the experimental value of $\tau_d$ and the value of $Q = 1.75 k_0$ at the center of ROI$_2$, we obtain $D_T \approx 2.3 \times 10^{-5}$ m$^2$/s. This value is lower than the bulk thermal diffusivity of Ir at room temperature, $D_T^{bulk} \approx 5 \times 10^{-5}$ m$^2$/s calculated from tabulated material parameters (see Supplementary Materials, Section S4), which is expected for a stack of ultrathin layers where interface scattering reduces thermal transport.

In addition to the surface displacement pattern that tracks the temperature profile, a spatially sinusoidal excitation at wave vector $\boldsymbol{Q}$ will launch counter-propagating SAWs which will also modulate the surface displacement and make the scattering intensity oscillate at the SAW frequency.[19] The weaker SAW oscillations within ROI$_2$ can be attributed to destructive interference between the SAWs generated via the LIPSS-like mechanism and those excited uniformly across the reciprocal space as described in Ref. [15]

**Conclusions**

We have shown that femtosecond laser irradiation below the damage threshold induces transient surface displacement patterns producing LIPSS-like features in EUV diffuse scattering. Just as in the case of permanent LIPSS, the root cause of the observed phenomenon is the interference of the radiation scattered by surface roughness with the incident beam resulting in a spatial modulation of the absorbed energy density. Unlike LIPSS, however, the patterns we observe do not involve any permanent changes to the sample surface; rather, they are produced by thermal expansion and disappear on the hundreds of picoseconds timescale as the spatial modulation of the temperature field is washed away by thermal diffusion.

Based on our observations, we believe that this effect is universal, occurring in any strongly absorbing material. The period of the transient LIPSS-like patterns is mainly controlled by the pump wavelength and varies between $\lambda_L/2$ and $\lambda_L$ depending on incidence angle and polarization; smaller periods can be achieved by using UV or EUV excitation. In ferro- and ferrimagnetic samples, transient magnetization textures[25] could be formed via the same mechanism.

Our approach provides a direct way to test electromagnetic scattering models such as Sipe's theory. While the formation of permanent LIPSS is a complicated nonlinear process and their parameters cannot be quantitatively predicted by the theory, diffuse EUV scattering by transient structures will enable a direct and reproducible comparison of experimental and theoretical maps of $A(Q)$. For example, the absence of distinct LIPSS-like features at high wave vectors[15] may indicate that the formation of high spatial frequency LIPSS[26] on metal samples is not explained by the pump radiation scattering.

Finally, our observations help disentangle different mechanisms responsible for the transient scattering signals reported in this work and in Ref. [15] Different time dynamics of the signals inside and outside the LIPSS-like feature shown in Fig. 3 indicate that two mechanisms must be in play here: whereas the absorption nonuniformity caused by light scattering from surface roughness is responsible for the LIPSS-like arcs, the uniform circular fringe pattern insensitive to the pump wavelength and polarization[15] must have a different origin. The mechanism based on the scattering of laser-generated longitudinal acoustic waves into SAWs[27] already suggested in Ref. [15] appears to be the most likely culprit.

**ACKNOWLEDGMENT**

D.K. and C.G. acknowledge funding by the Deutsche Forschungsgemeinschaft (DFG) Projects No. GU 535/9-1 and No. KS 62/3-1. The contribution by A.A.M. was supported by the Department of Energy, Office of Science, Office of Basic Energy Sciences, under award number DE-SC0019126. N.J. acknowledge financial support by the Agence Nationale de la Recherche, France, under grant agreement no. ANR-20-CE42-0012-01(MEDYNA) and no. ANR-21-CE09-0042 (HYPNOSE), by France 2030 government investment plan managed by the French National Research Agency under grant SPINCHARAC ANR-22-EXSP-0008. The team in Mainz acknowledges support by the DFG (Spin+X (A01, B01, A07), TRR 173 – 268565370 and Elasto-Q-Mat (A12) - 422213477).

We thank the FERMI free-electron laser facility for the beam time allocated under Proposal No. 2023406.

# REFERENCES

(1) Sipe, J. E.; Young, J. F.; Preston, J. S.; van Driel, H. M. Laser-Induced Periodic Surface Structure. I. Theory. *Phys. Rev. B* **1983**, *27* (2), 1141–1154. https://doi.org/10.1103/PhysRevB.27.1141.
(2) Young, J. F.; Preston, J. S.; Van Driel, H. M.; Sipe, J. E. Laser-Induced Periodic Surface Structure. II. Experiments on Ge, Si, Al, and Brass. *Phys. Rev. B* **1983**, *27* (2), 1155–1172. https://doi.org/10.1103/PhysRevB.27.1155.
(3) Bonse, J.; Hohm, S.; Kirner, S. V.; Rosenfeld, A.; Kruger, J. Laser-Induced Periodic Surface Structures— A Scientific Evergreen. *IEEE J. Sel. Top. Quantum Electron.* **2017**, *23* (3), 7581030. https://doi.org/10.1109/JSTQE.2016.2614183.
(4) Bonse, J.; Gräf, S. Maxwell Meets Marangoni—A Review of Theories on Laser-Induced Periodic Surface Structures. *Laser Photonics Rev.* **2020**, *14* (10), 2000215. https://doi.org/10.1002/lpor.202000215.
(5) Xu, S.; Zhang, Y.; Wang, T.; Zhang, L. Recent Developments of Femtosecond Laser Direct Writing for Meta-Optics. *Nanomaterials* **2023**, *13* (10), 1623. https://doi.org/10.3390/nano13101623.
(6) Wang, H.; Deng, D.; Zhai, Z.; Yao, Y. Laser-Processed Functional Surface Structures for Multi-Functional Applications-a Review. *J. Manuf. Process.* **2024**, *116*, 247–283. https://doi.org/10.1016/j.jmapro.2024.02.062.
(7) Nakhoul, A.; Colombier, J.-P. Beyond the Microscale: Advances in Surface Nanopatterning by Laser-Driven Self-Organization. *Laser Photonics Rev.* **2024**, *18* (5), 2300991. https://doi.org/10.1002/lpor.202300991.
(8) Vorobyev, A. Y.; Guo, C. Femtosecond Laser-Induced Periodic Surface Structure Formation on Tungsten. *J. Appl. Phys.* **2008**, *104* (6), 063523. https://doi.org//10.1063/1.2981072.
(9) Guosheng, Z.; Fauchet, P. M.; Siegman, A. E. Growth of Spontaneous Periodic Surface Structures on Solids during Laser Illumination. *Phys. Rev. B* **1982**, *26* (10), 5366–5381. https://doi.org/10.1103/PhysRevB.26.5366.
(10) Bonse, J.; Gräf, S. Ten Open Questions about Laser-Induced Periodic Surface Structures. *Nanomaterials* **2021**, *11* (12), 3326. https://doi.org/10.3390/nano11123326.
(11) Sikora, A.; Faucon, M.; Gemini, L.; Kling, R.; Mincuzzi, G. LIPSS and DLIP: From Hierarchical to Mutually Interacting, Homogeneous, Structuring. *Appl. Surf. Sci.* **2022**, *591*, 153230. https://doi.org/10.1016/j.apsusc.2022.153230.
(12) Bonse, J.; Sokolowski‐Tinten, K. Probing Laser‐Driven Structure Formation at Extreme Scales in Space and Time. *Laser Photonics Rev.* **2024**, *18* (5), 2300912. https://doi.org/10.1002/lpor.202300912.
(13) Kerber, N.; Ksenzov, D.; Freimuth, F.; Capotondi, F.; Pedersoli, E.; Lopez-Quintas, I.; Seng, B.; Cramer, J.; Litzius, K.; Lacour, D.; Zabel, H.; Mokrousov, Y.; Kläui, M.; Gutt, C. Faster Chiral versus Collinear Magnetic Order Recovery after Optical Excitation Revealed by Femtosecond XUV Scattering. *Nat. Commun.* **2020**, *11* (1), 6304. https://doi.org/10.1038/s41467-020-19613-z.
(14) Léveillé, C.; Burgos-Parra, E.; Sassi, Y.; Ajejas, F.; Chardonnet, V.; Pedersoli, E.; Capotondi, F.; De Ninno, G.; Maccherozzi, F.; Dhesi, S.; Burn, D. M.; Van Der Laan, G.; Latcham, O. S.; Shytov, A. V.; Kruglyak, V. V.; Jal, E.; Cros, V.; Chauleau, J.-Y.; Reyren, N.; Viret, M.; Jaouen, N. Ultrafast Time-Evolution of Chiral Néel Magnetic Domain Walls Probed by Circular Dichroism in x-Ray Resonant Magnetic Scattering. *Nat. Commun.* **2022**, *13* (1), 1412. https://doi.org/10.1038/s41467-022-28899-0.

(15) Capotondi, F.; Maznev, A.; Bencivenga, F.; Bonetti, S.; Fainozzi, D.; Fausti, D.; Foglia, L.; Gutt, C.; Jaouen, N.; Ksenzov, D.; Masciovecchio, C.; Nelson, K. A.; Nikolov, I.; Pancaldi, M.; Pedersoli, E.; Pfau, B.; Raimondi, L.; Romanelli, F.; Totani, R.; Trigo, M. Time-Domain Extreme Ultraviolet Diffuse Scattering Spectroscopy of Nanoscale Surface Phonons, 2025. https://arxiv.org/abs/2502.18445.

(16) Pedersoli, E.; Capotondi, F.; Cocco, D.; Zangrando, M.; Kaulich, B.; Menk, R. H.; Locatelli, A.; Mentes, T. O.; Spezzani, C.; Sandrin, G.; Bacescu, D. M.; Kiskinova, M.; Bajt, S.; Barthelmess, M.; Barty, A.; Schulz, J.; Gumprecht, L.; Chapman, H. N.; Nelson, A. J.; Frank, M.; Pivovaroff, M. J.; Woods, B. W.; Bogan, M. J.; Hajdu, J. Multipurpose Modular Experimental Station for the DiProI Beamline of Fermi@Elettra Free Electron Laser. *Rev. Sci. Instrum.* **2011**, *82* (4), 043711. https://doi.org/10.1063/1.3582155.

(17) Bencivenga, F.; Mincigrucci, R.; Capotondi, F.; Foglia, L.; Naumenko, D.; Maznev, A. A.; Pedersoli, E.; Simoncig, A.; Caporaletti, F.; Chiloyan, V.; Cucini, R.; Dallari, F.; Duncan, R. A.; Frazer, T. D.; Gaio, G.; Gessini, A.; Giannessi, L.; Huberman, S.; Kapteyn, H.; Knobloch, J.; Kurdi, G.; Mahne, N.; Manfredda, M.; Martinelli, A.; Murnane, M.; Principi, E.; Raimondi, L.; Spampinati, S.; Spezzani, C.; Trovò, M.; Zangrando, M.; Chen, G.; Monaco, G.; Nelson, K. A.; Masciovecchio, C. Nanoscale Transient Gratings Excited and Probed by Extreme Ultraviolet Femtosecond Pulses. *Sci. Adv.* **2019**, *5* (7), eaaw5805. https://doi.org/10.1126/sciadv.aaw5805.

(18) Eichler, H. J.; Günter, P.; Pohl, D. W. *Laser-Induced Dynamic Gratings*; Tamir, T., Series Ed.; Springer Series in Optical Sciences; Springer Berlin Heidelberg: Berlin, Heidelberg, 1986; Vol. 50. https://doi.org/10.1007/978-3-540-39662-8.

(19) Rogers, J. A.; Maznev, A. A.; Banet, M. J.; Nelson, K. A. Optical Generation and Characterization of Acoustic Waves in Thin Films: Fundamentals and Applications. *Annu. Rev. Mater. Sci.* **2000**, *30* (1), 117–157. https://doi.org/10.1146/annurev.matsci.30.1.117.

(20) Bonse, J.; Munz, M.; Sturm, H. Structure Formation on the Surface of Indium Phosphide Irradiated by Femtosecond Laser Pulses. *J. Appl. Phys.* **2005**, *97* (1), 013538. https://doi.org/10.1063/1.1827919.

(21) Käding, O. W.; Skurk, H.; Maznev, A. A.; Matthias, E. Transient Thermal Gratings at Surfaces for Thermal Characterization of Bulk Materials and Thin Films. *Appl. Phys. A* **1995**, *61* (3), 253–261. https://doi.org/10.1007/BF01538190.

(22) Goodman, J. W. *Introduction to Fourier Optics*, 3rd ed.; Greenwood Village, 2005.

(23) Maznev, A. A.; Nelson, K. A.; Rogers, J. A. Optical Heterodyne Detection of Laser-Induced Gratings. *Opt. Lett.* **1998**, *23* (16), 1319. https://doi.org/10.1364/OL.23.001319.

(24) Thomsen, C.; Grahn, H. T.; Maris, H. J.; Tauc, J. Surface Generation and Detection of Phonons by Picosecond Light Pulses. *Phys. Rev. B* **1986**, *34* (6), 4129–4138. https://doi.org/10.1103/PhysRevB.34.4129.

(25) Ksenzov, D.; Maznev, A. A.; Unikandanunni, V.; Bencivenga, F.; Capotondi, F.; Caretta, A.; Foglia, L.; Malvestuto, M.; Masciovecchio, C.; Mincigrucci, R.; Nelson, K. A.; Pancaldi, M.; Pedersoli, E.; Randolph, L.; Rahmann, H.; Urazhdin, S.; Bonetti, S.; Gutt, C. Nanoscale Transient Magnetization Gratings Created and Probed by Femtosecond Extreme Ultraviolet Pulses. *Nano Lett.* **2021**, *21* (7), 2905–2911. https://doi.org/10.1021/acs.nanolett.0c05083.

(26) Bonse, J.; Kirner, S. V.; Krüger, J. Laser-Induced Periodic Surface Structures (LIPSS). In *Handbook of Laser Micro- and Nano-Engineering*; Sugioka, K., Ed.; Springer: Cham, 2021; pp 879–936. https://doi.org/10.1007/978-3-030-63647-0_17.

(27) Maznev, A. A. Boundary Scattering of Phonons: Specularity of a Randomly Rough Surface in the Small-Perturbation Limit. *Phys. Rev. B* **2015**, *91* (13), 134306. https://doi.org/10.1103/PhysRevB.91.134306.

# Transient laser-induced periodic surface structures revealed by time-resolved EUV diffuse scattering

## *Supplemental Material*

D. Ksenzov,[1,*,†] F. Capotondi,[2,*] A. A. Maznev,[3,*] F. Bencivenga,[2] D. Engel,[4] D. Fausti,[2,5,6] L. Foglia,[2] R. Gruber,[7] N. Jaouen,[8,9] M. Kläui,[7,10] I. Nikolov,[2] M. Pancaldi,[2,8] E. Pedersoli,[2] B. Pfau,[4] C. Gutt[1]

[1]Universität Siegen, Walter-Flex-Straße 3, 57072 Siegen, Germany.

[2]Elettra Sincrotrone Trieste, Strada Statale 14, km 163.5, 34149 Basovizza, TS, Italy.

[3]Department of Chemistry, Massachusetts Institute of Technology Cambridge, Massachusetts 02139, USA.

[4]Max Born Institute, Max-Born-Straße 2A, 12489 Berlin, Germany.

[5]Department of Physics, Università degli Studi di Trieste, 34127, Trieste, Italy

[6]Department of Physics, University of Erlangen-Nürnberg, 91058, Erlangen, Germany

[7]Institut für Physik, Johannes Gutenberg-Universität Mainz, 55099, Mainz, Germany

[8]Department of Molecular Sciences and Nanosystems, Ca' Foscari University of Venice, Venice, Italy

[9]Synchrotron SOLEIL, Saint-Aubin, Boite Postale 48, 91192, Gif-sur-Yvette Cedex, France.

[10]Graduate School of Excellence Materials Science in Mainz, 55128, Mainz, Germany

* These authors contributed equally to this work.

† Contact author: dmitriy.ksenzov@uni-siegen.de

**S1. Comparison of differential DES intensity images at various probe wavelengths**

To investigate the robustness of the observed ultrafast dynamics in the multilayer stack [Ir (6.6 nm)/CoFeB (0.7 nm)/MgO (2.0 nm)]x15, we performed differential diffusion EUV scattering (DES) measurements using probe pulses at three different wavelengths: 22.5 nm, 20.8 nm, and 19.3 nm, while maintaining a fixed pump wavelength of 790 nm. These measurements were conducted under identical experimental conditions, including a fixed detector-to-sample distance of 140 mm, the same number of shots accumulated per frame, and the incident pump fluence. This ensures that any observed differences in the scattering patterns arise solely from the variation in probe wavelength, rather than from changes in experimental setup.

Figure S1 presents the differential DES intensity images recorded at a fixed pump–probe delay of $t = 130$ ps. These images reveal transient LIPSS-like features formed in the multilayer structure following excitation with 790-nm p-polarized pump pulses. The periodic surface modulation is clearly visible across all three probe wavelengths, demonstrating that the spatial characteristics of the excited state are consistently captured, independent of the EUV probe wavelength.

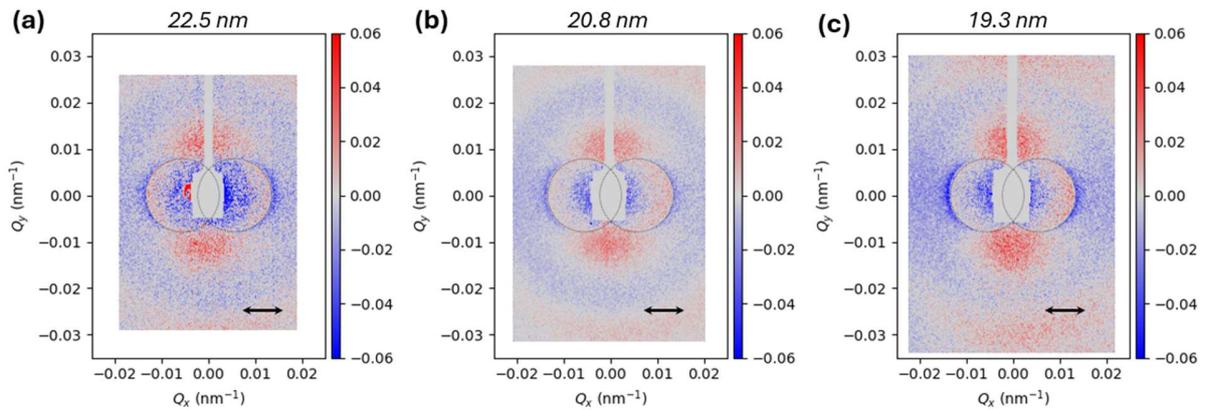

**Figure S1.** Differential DES intensity patterns recorded at a fixed pump–probe delay of $t = 130$ ps, demonstrating the LIPSS-like futures in the multilayer stack [Ir/CoFeB/MgO]x15 under 790-nm pump excitation with p-polarized light, recorded using EUV probe pulses at wavelengths of **(a)** 22.5 nm, **(b)** 20.8 nm, and **(c)** 19.3 nm. Dotted lines indicate the positions of LIPSS wavevectors, calculated based on the scattering geometry and an incidence angle of 45°.

Moveover, the temporal evolution of the DES signal remains qualitatively unchanged across the different probe wavelengths. This consistency confirms that the dynamics observed within the selected Q-range are intrinsic to the sample response and not an artifact of the probing conditions. The ability to reproduce the same structural features using multiple EUV wavelengths strengthens the reliability of the DES technique for capturing picosecond surface and interface dynamics in complex multilayer systems.

### S2. Transient LIPSS-like features in metallic thin films

To demonstrate that the formation of transient LIPSS-like features is not limited to multilayer systems, we also investigated their appearance in single-layer metallic thin films. Specifically, we studied platinum (Pt), cobalt (Co), tantalum (Ta), and titanium (Ti) films of varying thicknesses, all of them grown on Si (001) substrate by magneto sputtering deposition. Each sample was excited by 395-nm femtosecond laser pulses with s-polarization. The differential DES signal was recorded at a fixed pump–probe delay of $t = 50$ ps, corresponding to the peak contrast of surface modulation observed in the scattering patterns.

Figure S2 presents the differential DES intensity image, revealing arc-like features oriented perpendicular to the laser polarization direction. This orientation is consistent with typical LIPSS behavior under s-polarized excitation. Black arrows indicate the direction of laser polarization, while gray arrows denote the orientation and position of the transient surface modulations. Gray areas mark the location of beamstops used during the measurement.

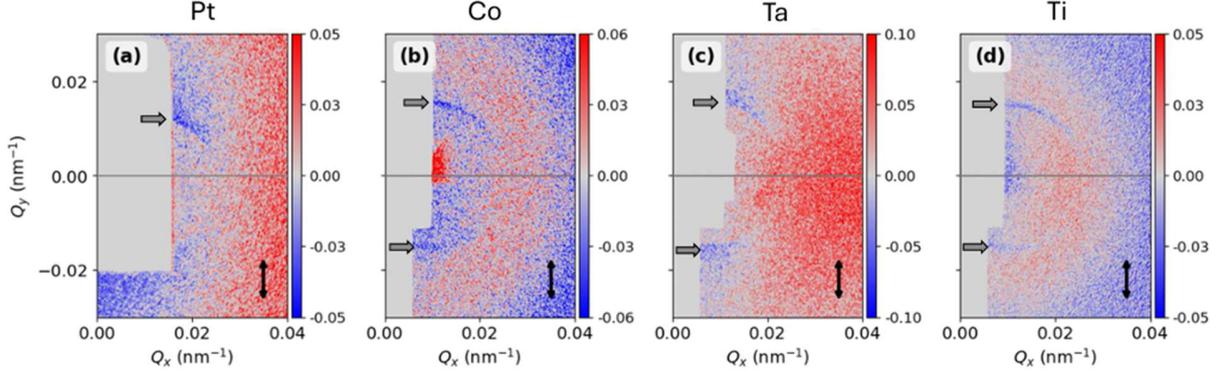

**Figure S2**. Differential DES intensity patterns recorded at a fixed pump–probe delay of $t = 50$ ps, showing the LIPSS-like features in thin metallic films under 395-nm pump excitation with s-polarized light. Each panel corresponds to a different film: **(a)** 50-nm Pt, **(b)** 75-nm Co, **(c)** 75-nm Ta, and **(d)** 100-nm Ti. Black arrows indicate the laser polarization direction; gray arrows mark the position and orientation of the LIPSS-like features. The gray areas mark the location of the beamstops.

### S3. Computation of the modulation coefficient $A$

The modulation coefficient $A(\beta_x)$ is related to the coefficient $P_c$ from Guosheng et al. [S2] as follows,

$$A = -\frac{P_c}{k_0 h}. \tag{S3.1}$$

We calculated $P_c (\beta_x)$ using Eq. (17) of Ref. [S2] after correcting the following typos: the term $(\alpha_s^1 + \alpha_s^{-1})$ should be replaced by its complex conjugate, and the sign of the entire expression should be flipped. (We verified that after these corrections the equation reproduces the graphical results presented in Ref. [S2].) The calculations were performed using the complex refractive index of Iridium at 790 nm $n = 3.4 + i6.2$ [S3] for an incidence angle of 45° and p-polarization.

**S4. Thermal diffusivity**

The thermal diffusivity of a material is related to its thermophysical constants through

$$D_T = \frac{\kappa}{\rho c} \qquad (S4.1)$$

where $\kappa$ is the thermal conductivity, $\rho$ is the mass density, and $c$ is the specific heat capacity at constant pressure. The material parameters used to calculate the bulk diffusivity of Ir at room temperature are listed in Table S1.

**Table S1.** Thermophysical properties of Ir at room temperature

| Parameter | Symbol | Value | Source |
|---|---|---|---|
| Thermal conductivity | $\rho$ | 150 W/(m·K) | [S4] |
| Density | $\rho$ | 22,650 kg/m$^3$ | [S5] |
| Specific heat capacity | $c$ | 131 J/(kg·K) | [S6] |
| Thermal diffusion (calculated) | $D_T = \frac{\kappa}{\rho c_p}$ | 5.1×10$^{-5}$ m$^2$/s | This work |

**References**

[S1] Sipe, J. E.; Young, J. F.; Preston, J. S.; van Driel, H. M. Laser-Induced Periodic Surface Structure. I. Theory. *Phys. Rev. B* 1983, *27*, 1141–1154.

[S2] Guosheng, Z., G.; Fauchet, P. M.; Siegman, A. E. Growth of Spontaneous Periodic Surface Structures on Solids during Laser Illumination. *Phys. Rev. B* **1982**, *26*, 5366–5381.

[S3] Schmitt, P.; Felde, N.; Döhring, T.; Stollenwerk, M.; Uschmann, I.; Hanemann, K.; Siegler, M.; Klemm, G.; Gratzke, N.; Tünnermann, A.; Schwinde, S.; Schröder, S.; Szeghalmi, A. Optical,

Structural, and Functional Properties of Highly Reflective and Stable Iridium Mirror Coatings for Infrared Applications. *Opt. Mater. Express* **2022**, *12*, 545–559.

[S4] Ho, C. Y.; Powell, R. W.; Liley, P. E. Thermal Conductivity of the Elements. *J. Phys. Chem. Ref. Data* **1972**, *1* (2), 279–421. https://doi.org/10.1063/1.3253100

[S5] *Landolt–Börnstein – New Series*, Volume IV, Phase Equilibria, Crystallographic and Thermodynamic Data of Binary Alloys; Springer: Berlin, 1990.

[S6] Furukawa, G. T.; Reilly, M. L.; Gallagher, J. S. Critical Analysis of Heat-Capacity Data and Evaluation of Thermodynamic Properties of Ruthenium, Rhodium, Palladium, Iridium, and Platinum from 0 to 300 K. A Survey of the Literature Data on Osmium. *J. Phys. Chem. Ref. Data* **1974**, *3* (1), 163–209. https://doi.org/10.1063/1.3253137